\documentclass[a4paper]{jpconf}
\usepackage{bbm}
\usepackage{amsmath}
\usepackage{amsfonts}
\usepackage{amssymb}
\usepackage{amsthm}
\usepackage{slashed}
\newcommand{\llang}{\langle \! \langle}
\newcommand{\rrang}{\rangle \! \rangle}

\begin{document}
\title{Quantum Gravity and Cosmology: an intimate interplay}

\author{Mairi Sakellariadou}

\address{Theoretical Particle Physics and Cosmology Group, Department of Physics, King's College London, University of London, Strand, London, WC2R 2LS, U.K.}

\ead{mairi.sakellariadou@kcl.ac.uk}

\begin{abstract}
I will briefly discuss three cosmological models built upon three distinct quantum gravity proposals. I will first highlight the cosmological r\^ole of a vector field in the framework of a string/brane cosmological model. I will then present the resolution of the big bang singularity and the occurrence of an early era of accelerated expansion of a geometric origin, in the framework of 
group field theory condensate cosmology. I will then summarise results from an extended gravitational model based on non-commutative 
spectral geometry, a model that offers a purely geometric explanation for the standard model of particle physics.

\end{abstract}

\section{Introduction}

Most of the fundamental questions in cosmology - which for a long time belonged to the realm of philosophy -  can now be addressed within a physical theory and tested against increasingly accurate observational data. We  live the golden age of cosmology. Still, fundamental puzzles remain. The most obvious is {\sl what, if anything, happened before the big bang?} Another is {\sl what is the universe made of?}

Standard cosmology is formulated within the framework of Einstein's general theory of relativity. Notwithstanding, general relativity is not adequate to explain the earliest stages of cosmic existence, and cannot provide an explanation for the big bang itself. The reason is that  quantum mechanics must have played a r\^ole at sufficiently early times. 
The majority of early universe cosmological models are studied at a semi-classical level, hence they leave basic questions such as the beginning of the universe, and the origin of inflation (or of alternative models) unanswered. Modern early universe cosmology is  in the need of a rigorous underpinning in quantum gravity. More precisely, quantum gravity offers the appropriate framework  to address fundamental questions such as the origin of the space-time continuum and its effective dynamics, and the pictures of the earliest moments of the universe which emerge. These  pictures  must include an approximately homogeneous background and superimposed perturbations. In return, cosmological data represent the best (if not the only) chance for testing quantum gravity, thus guiding the formulation of the complete fundamental theory. 

Quantum gravity proposals belong to two classes, they are either {\sl top-down} or {\sl bottom-up}. In the former class belong approaches like string/M-theory or a variant of non-perturbative approach to quantum gravity, like loop quantum gravity or group field theory. In the latter one belong the proposal of non-commutative (spectral) geometry, and that of asymptotic safety.  
String theory may provide the underlying theory required  to resolve questions such as the initial big bang singularity, the origin of an early inflationary era or of an alternative to inflation mechanism, and explain  space-time dimensionality~\cite{bv,md}.
Group field theory is a discrete quantum gravity  approach which relies on pre-gerometric structures.  Its hardest challenge is to retrieve, within an appropriate (semi-classical) limit, the geometry of continuum space-times and the corresponding dynamics as given by (classical) general relativity.
A main tenet of non-commutative geometry is that above the Planck scale the whole concept of geometry breaks down and space-time is replaced by a non-commutative manifold. In this approach, below but close to the Planck energy scale, space-time is the product of a Riemannian spin manifold with a finite non-commutative space.

In what follows, I will highlight examples which show the intimate relation between the search for a theory of quantum gravity and the urge to find the appropriate theoretical framework to address questions about the origin and very early moments of our universe and unweave its fabric. I will first highlight a cosmological model based on a microscopic string theory proposal, called the {\sl D-material universe}. I will then present the effective dynamics of a cosmological model, called the {\sl group field theory condensate cosmology}, derived from group field theory.

\section{A string theory inspired cosmological model: D-material universe}
D-material universe~\cite{nmetal} is a modified gravity model involving a fundamental vector field; it may appear as the low-energy limit of certain brane theories.

Consider a compactified (3+1)-brane propagating in a higher dimensional bulk, populated by point-like D0-branes, called D-particles. Since a brane
is, by definition, the collection of end-points of open strings, this picture
implies that open strings propagate in a medium of D-particles. Thus, 
brane-puncturing massive D-particles can be captured by electrically 
neutral matter (to be consistent with the fact that we are
living in an electrically neutral universe) open strings. As a result, 
D-particles recoil and their velocity will locally break Lorentz invariance.
The emerged vector-like excitations can lead to an early era of geometric inflation and contribute to large scale structure, enhancing dark matter perturbations~\cite{nmetal,furqaan-nick-mairi}. This is an example of a bi-metric theory, there is the sigma model background metric related to the Einstein-frame metric, and a metric describing the distortion of space-time surrounding D-particles~\cite{nick-mairi}. 

The interaction between stringy matter (living in a brane-world of three longitudinal large dimensions) and a medium of recoiling D-particles, is given using the Dirac-Born-Infeld (DBI) action:
\begin{eqnarray}
\label{DBI}
S_{\rm eff ~ 4D}&=&\int {\rm d}^4 x\Big[-{\sqrt{-g}\over 4}e^{-2\phi}{\cal G}_{\mu\nu}{\cal G}^{\mu\nu}-{T_3\over g_{{\rm s}0}}e^{-\phi}\sqrt{-{\rm det}(g+2\pi\alpha' F)}(1-\alpha R(g))\nonumber\\
&&\ \ \ \ \ \ \ \ \ -\sqrt{-g}{e^{-2\phi}\over \kappa_0^2}\tilde\Lambda + \sqrt{-g}{e^{-2\phi}\over \kappa_0^2}R(g)+{\cal O}\Big((\partial\phi)^2\Big)\Big]+S_{\rm m}~,
\end{eqnarray} 
where ${\cal G}_{\mu\nu}$ is a flux gauge field, $T_3$ denotes the brane tension, $\phi$ is the dilaton field assumed to be constant, $g_{\rm s}=g_{{\rm s}0}e^\phi$ is the string coupling, which  is just $g_{{\rm s}0}$ for a constant dilaton, $g$ stands for the determinant of the gravitational field, 
$\tilde\Lambda$ is the cosmological constant, $\kappa^{-2}$ is the four-dimensional bulk induced gravitational constant $\kappa^{-2}=\big(V^{(6)}/ g_{{\rm s}0}^2\big)M_{\rm s}^2$ with $M_{\rm s}$ the string scale $M_{\rm s}=1/\sqrt{\alpha'}$. The vector field $A_\mu$ denotes the recoil velocity excitation during the string-matter with D-particle interactions; its field strength is
\begin{equation}
F_{\mu\nu}\equiv \partial_\mu A_\nu-\partial_\nu A_\mu~.
\end{equation}
The vector field  $A_\mu$ satisfies the constraint
\begin{equation}
\label{constraint}
A_\mu A_\nu g^{\mu\nu}=-{1\over\alpha'}~.
\end{equation}
In the low-energy weak approximation, one can expand the four-dimensional DBI action (\ref{DBI}) and get, in the Einstein frame  $g_{\mu\nu}^{\rm E}=e^{-2\phi}g_{\mu\nu}$, the effective action
\begin{eqnarray}
\label{effS}
S_{\rm eff~~4D}^{\rm E}&=&\int{\rm d}^4 x \sqrt{-g}
\Big[-{T_3e^{3\phi_0}\over g_{{\rm s}0}}
- {\tilde\Lambda e^{2\phi_0}\over  \kappa_0^2}
-{\tilde F^2\over 4}(1-\alpha e^{-2\phi_0} R)
+\Big({\alpha T_3 e^\phi_0\over g_{{\rm s}0}}+{1\over \kappa_0^2}
\Big)R
\nonumber\\
&&+\lambda \Big(\tilde A_\mu \tilde A^\mu +{4\pi^2\alpha' T_3 e^{3\phi_0}\over g_{{\rm s}0}}
\Big)\cdots \Big]+S_{\rm m}~,
\end{eqnarray}
where $\tilde F_{\mu\nu}$ is the Maxwell field strength for the redefined vector field
\begin{equation}
{\tilde A}_\mu\equiv \Big({T_3(2\pi\alpha')^2\over g_{{\rm s}0}} e^{3\phi_0}
\Big) ^{1/2} A_\mu~,
\end{equation}
and $\alpha=\alpha'\zeta(2)=\alpha'\pi^2/6$.
The last term in (\ref{effS}) implements the constraint
(\ref{constraint}) rewritten for the redefined vector field ${\tilde A}_\mu$ as
\begin{equation}
\tilde A_\mu \tilde A_\nu g^{\mu\nu}=-{4\pi^2\alpha' T_3 e^{3\phi_0}\over g_{{\rm s}0}}~;
\end{equation} 
$\lambda$ is a Lagrange multiplier.

Having the action in the low-energy weak-field approximation, therefore valid at late times, one can derive the graviton, vector field and dilaton equations of motion. The vector field equation of motion fixes the background value of the Lagrange multiplier field:
\begin{equation}
\langle \lambda(x)\rangle ={e^{-3\phi_0}g_{{\rm s}0}\over 8\pi^2\alpha'|T_3|}
\tilde A_\mu [\tilde F_{\nu\mu}(1-\alpha e^{-2\phi_0}R)]^{;\nu}~,
\end{equation}
while considering galactic scales, the dilaton equation of motion fixes the value of the  cosmological constant:
\begin{equation}
\Lambda_0\simeq - {T_3\over 2 g_{{\rm s}0}} e^{3\phi_0}~, 
\end{equation}
which is negative for a positive brane tension $T_3$. Note that such anti-de Sitter type terms cancel against dilaton independent contributions to the brane vacuum energy, thus leading to a small positive cosmological constant
\begin{equation}
\Lambda^{\rm vac}\equiv \Lambda_0+{1\over 8}\langle {\cal G}_{\mu\nu}
{\cal G}^{\mu\nu}\rangle +\cdots >0~,
\end{equation}
during the galactic era. 

At sufficiently late scales, so that the effective action (\ref{effS}) is valid, we will analyse gravitational lensing phenomenology. We consider a static spherically symmetric background
\begin{equation}
g_{\alpha\beta}{\rm d}x^\alpha{\rm d}x^\beta
=-e^{\nu(\sqrt{x^2+y^2+z^2})}{\rm d}t^2
+ e^{\zeta(\sqrt{x^2+y^2+z^2})}a^2(t)({\rm d}x^2+{\rm d}y^2+{\rm d}z^2)~.
\end{equation} 
Interactions of D-particles with open strings lead to recoil fluctuations of
the former, corresponding to world-sheet deformations of gauge fields, as
\begin{equation}
\label{Ai}
 A_i({\vec x},t)\simeq {1\over \alpha'}g_{ij}({\vec x},t)u^j
\Big(t{a(t_{\rm c})^2\over a(t)^2}-t_{\rm c}
\Big)~,~t>t_{\rm c}~,
\end{equation}
with $t_{\rm c}$ denoting the collision time, which is of the same order of magnitude as the Friedmann-Lema\-{i}tre-Robinson-Walker 
cosmic time, hence
\begin{equation}
\label{tc}
t_{\rm c}\sim{2\over H_0[1+(1+z)^2]}~.
\end{equation}
for a galaxy at redshift $z$.  The time of observation $t$ is considered as the present time $t_0$.
Using Eq.~(\ref{Ai}) with the constraint 
(\ref{constraint}) and the definition (\ref{tc}) with $a(t_{\rm c})=a_0/(1+z)$, 
one finds that the {\sl electric} type field strength components $F_{ti}$,
\begin{equation}
F_{ti}({\vec x},t)\sim{1\over \alpha'}g_{ij}({\vec x},t)u^j\Big[
{1-3(1+z)^2\over(1+z)^2(1+(1+z)^2)}
\Big]~,
\end{equation}
associated with linear recoil momentum excitations, are much bigger than the {\sl magnetic} type field strength components $F_{ij}$, corresponding to non-zero angular momentum of recoiling D-particles. Hence the latter may be neglected.

Considering late eras of the cosmological evolution, D-particles are assumed to have recoil velocities following gaussian stochastic statistics:
\begin{equation}
\llang u^m u^n\rrang=\sigma_0^2(t)\delta^{mn}~,~
\llang u^m\rrang=0~,~
\sigma_0^2(t)=[a(t)]^{-3}|\beta|~,
\end{equation}
 so that macroscopically Lorentz invariance is maintained. 
The statistical fluctuations are proportional to the cosmic density of defects, with constant of proportionality the statistical variance $|\beta|$ of the recoil velocity:
\begin{equation}
|\beta|\sim {1\over 3}{n_{\rm D}\over n_\gamma}{\xi^2|p_i^{\rm phys}|^2\over M_{\rm s}^2} g_{{\rm s}0}^2~,
\end{equation}
where
$n_{\rm D}$, $n_\gamma$ denote respectively the number density of D-particles and radiation, $\xi$ is a space-time local constant fudge factor, given by the microscopic theory and in general smaller than unity.
Assuming scattering of D-particles is mainly with cosmic photons, 
$p_i$ is the presently observed average energy density of cosmic microwave background (CMB) photons:
\begin{equation}
\sqrt{|p_i^{\rm phys CMB}|^2}\sim 7\times 10^{-4} {\rm eV}~.
\end{equation}
The string scale is $10^4 {\rm GeV}\leq M_{\rm s}\leq 10^{18} {\rm GeV}$.

One then explores the parameter space of $|\beta|$, so that D-particles can play the r\^ole of dark matter candidates, while at the same time they can provide the seed for large scale structure. A detailed analysis in Ref.~\cite{nmetal}, combining lensing data with an estimate of the required densities so that the D-matter recoil-velocity fluid can mimic dark matter in galaxies, in the sense that its contribution to the energy density is of the same order as the mass density of a galaxy, led to the constraint
\begin{equation}
10^{-58} M_{\rm Pl}^4\leq T_3\leq 10^{-26} M_{\rm Pl}^4~.
\end{equation}
At this point, let us emphasise that in this model we do not suggest a mechanism to replace dark matter, since we have for instance neutrinos, instead in our scenario the r\^ole of dark matter is enhanced by the vector field. 

Moreover, neutrinos may appear as dark matter candidates that could be captured by D-particles. After the capture by the D-particle defect, the emerging stringy matter excitation may have a different flavor than its original one. Thus, D-particle populations in galaxies may act as a {\sl medium} inducing flavor oscillations.
Considering only $\nu_e\leftrightarrow \nu_\mu$ oscillations and computing the average of the neutrino stress tensor with respect to the flavor vacuum, the obtained~\cite{nick-mairi} extra time-dependent dark matter energy contribution is
\begin{equation}
\Omega_\Lambda^{\nu_{\rm mixing}}\sim 0.24~.
\end{equation}
D-particles may also induce a successful inflationary mechanism through condensation of their recoil velocity field. Such a mechanism is compatible with the latest Planck data provided the condensate fields are large ($\sigma(t)\gg 1$), namely we have dense D-particle populations in the early universe, which one expects to be sufficiently diluted today.
This regime is appropriate for low string mass scales with respect to the  hubble  inflationary scale:
\begin{equation}
M_{\rm s}\ll H_{\rm I}\sim 10^{-5} M_{\rm Pl}\ll M_{\rm Pl}~,
\end{equation}
using current Planck data.

Using finite temperature formalism (Euclidean time) to account for the Hawking-Gibbons temperature (associated with an observer-dependent horizon) of a de Sitter space-time, one gets~\cite{nmetal} the Euclidean DBI action and then through analytic continuation one recovers a
Minkowski space.
The Euclidean DBI action reads 
\begin{equation} 
	S_{\rm eff~4D} \simeq \int d^4x \sqrt{\tilde g} \,\frac{1}{\kappa_0^2} \left[ R(\tilde g) + \frac{1}{2} \,\partial_\mu \varphi \,\partial^\mu \varphi - \frac{\kappa_0^2}{4} \,{\mathcal G}_{\mu\nu} {\mathcal G}^{\mu\nu} - \frac{e^{-\sqrt{\frac{2}{3}} \,\varphi}}{\alpha} - \left( \tilde \Lambda - \frac{1}{\alpha} \right) e^{-2 \sqrt{\frac{2}{3}} \,\varphi} \right]~,
\end{equation}
where we have redefined the metric as
\begin{equation}
\tilde g_{\mu\nu}=(1+\sigma)g_{\mu\nu}~,
\end{equation}
and $\varphi$ denotes a canonically normalised scalar field:
\begin{equation}
\varphi={\sqrt{3\over 2}}\ln(1+\sigma(t))~.
\end{equation} 
Assuming the flux field condensates into a constant one, which contributes to the vacuum energy as 
\begin{equation} 
	\frac{1}{4} \llang {\mathcal G}_{\mu\nu} \,{\mathcal G}^{\mu\nu} \rrang \equiv {\mathcal D} ~,
\end{equation}
the {\sl Euclideanised} effective potential for the canonically normalised scalar field, reads
\begin{equation} 
	V^{\mathcal E} = - \kappa_0^2 \,\mathcal D - \frac{e^{-\sqrt{\frac{2}{3}} \,\varphi}}{\alpha} - \left( \tilde \Lambda - \frac{1}{\alpha} \right) e^{-2 \sqrt{\frac{2}{3}} \,\varphi}~.
\end{equation}
Performing analytic continuation back to Minkowski, teh potential is approximated by
\begin{align}\label{v}
	V(\varphi) \simeq \kappa_0^2 \,{\mathcal D} + \left( \tilde \Lambda - \frac{1}{\alpha} \right) e^{-2\sqrt{\frac{2}{3}} \,\tilde\varphi} + i \,\frac{e^{-\sqrt{\frac{2}{3}} \,\tilde\varphi}}{\alpha} ~,
\end{align}
with $\tilde\varphi$ real. The imaginary part in the effective potential above denotes an instability; it has the following important physical meaning. The field rolls down, the condensate becomes small, the imaginary part disappears, and consequently one can expand the square-root of the DBI action and recover the effective action valid at late eras, which we have used earlier to study gravitational lensing and growth of matter perturbations.

The real part of the effective potential (\ref{v}) is~\cite{nmetal} 
\begin{equation} \label{star}
	{\rm Re} \,V(\tilde \varphi ) = \tilde {\mathcal D} + \left( \tilde \Lambda - \frac{1}{\alpha} \right) e^{-2 \sqrt{\frac{2}{3}} \,\tilde\varphi}~, \quad \tilde{\mathcal D} \equiv \kappa_0^2 \,{\mathcal D}~,
\end{equation}
with $(\tilde \Lambda - 1/\alpha)$ negative relative to  $\tilde {\mathcal D}$.
The above potential (\ref{star})
is of Starobinsky type, provided one can tune the flux-field condensate to be $\tilde {\mathcal D} > 0$ and such that the minimum of the potential occurs for the field value $\tilde \varphi = 0$ and corresponds to zero potential. 

Fixing the spectral index to $n_{\rm s}=0.965$ we get $N=57.7$ e-folds, leading to the slow-roll parameters $\epsilon, \xi, \eta$~\cite{nmetal}
\begin{equation} 
	\epsilon\simeq 5.6 \times 10^{-5} \ll 1 ~, \qquad \eta \simeq - 1.7 \times 10^{-2} \ll 1 ~, \qquad \xi \simeq 3.0 \times 10^{-4} \ll 1~,
\end{equation}
consistent with current CMB data.
The CMB constraint on $V/\epsilon$ leads to
\begin{equation} 
	{\mathcal D} \simeq 3.2 \times 10^{-11} \;M_{\rm Pl}^4 \qquad \Leftrightarrow \qquad \tilde {\mathcal D} \simeq 3.2 \times 10^{-11} \;M_{\rm Pl}^2~.
\end{equation}
It is worth noting that it is the gauge field flux condensate   ${\cal G}_{\mu\nu}{\cal G}^{\mu\nu}$             that induces a de Sitter phase (positive, almost constant, vacuum energy), and hence inflation, but it is the recoiling D-particles velocity vector field that induces a slowly rolling scalar degree of freedom that allows exit of inflation.

\section{A non-perturbative approach to QG: Group Field Theory Condensate Cosmology}
\subsection{The setup}

Group Field Theory (GFT)~\cite{GFT,GFTReview}, a second quantisation formulation of loop quantum gravity, states that spacetime and geometry emerge through an effective description of the collective behaviour of pre-geometric degrees of freedom. This approach conjectures that the  continuous macroscopic space-time we are familiar with, emerges through a condensation of bosonic GFT quanta, each representing an atom of space.
Within this context, classical gravitational dynamics is recovered through an effective description of the GFT condensate dynamics. 

A GFT model is defined by a choice of three ingredients: a group that plays the r\^ole of the  local gauge group of gravity, a fundamental building block of space, and a field with its corresponding action encoding the dynamics. The main choices so far are respectively, the gauge group SU(2) of Ashtekar-Barbero gravity, a quantum tetrahedron created or annihilated by second quantised field operators, and a complex scalar field with an action given by a kinetic quadratic term and 
a sum of interaction polynomials weighted by corresponding coupling constants. The interaction terms implement the way the quanta of geometry are {\sl glued} with each other.  

The basic idea of GFT is that all data encoded in the fields is only of combinatorial and algebraic nature. Hence, GFT is a background-independent field theory. In the simplest case, the data, storing geometric information, is made of group elements attached to the edges of a graph.  The data correspond to the holonomies of a connection and the fluxes of a triad field, which are the fundamental variables of loop quantum gravity. The group elements can be seen as parallel transports of a gravitational connection along the edges of a graph. 

Being interested in building a four-dimensional quantum gravity, the complex scalar field $\varphi$ lives on $d=4$ copies of the $G=$SU(2) group: $\varphi(g_I): G^d\rightarrow \mathbbm{C}$, with $I=1,\cdots, d$, where the group elements $g_I$ correspond to parallel transports 
on the gravitational connection along a link. In addition, one imposes invariance of the field $\varphi$ under the simultaneous right
multiplication of all $g_I\in G$ by the same group element $h$, for all $h\in G$: $\varphi(g_1h,\cdots,g_dh)=\varphi(g_1,\cdots,g_d), \forall h\in G$.
The classical action can be written as
\begin{equation}
S[\varphi,\bar{\varphi}]=\int_G(dg)^d\int_G(dg')^d\bar{\varphi}(g_I)\mathcal{K}(g_I,g'_I)\varphi(g_I)+\mathcal{V}[\varphi,\bar{\varphi}]~,
\label{cl-ac}\end{equation}
where $\mathcal{K}$ is the kinetic kernel and $\mathcal{V}$ represents the interaction term; the former is typically local whereas the latter is in general a nonlinear and nonlocal convolution of $\varphi$ with itself. The above action leads to the classical equation of motion 
\begin{equation}
\int_G(dg')^d\mathcal{K}(g'_I,g_I)\varphi(g_I)+\frac{\delta \mathcal{V}}{\delta\bar{\varphi}(g_I)}=0~,
\end{equation}
for the GFT field.
In the quantum version of the theory, the dynamics is obtained from the partition function
\begin{equation}
Z=\int[\mathcal{D}\varphi][\mathcal{D}\bar{\varphi}]e^{-S[\varphi,\bar{\varphi}]}~.
\end{equation}
A no-space state, which therefore does not include any topological or quantum geometric information, is defined by the Fock vacuum:
\begin{equation}
\hat{\varphi}(g_I)|\emptyset\rangle=0~.
\end{equation}
The excitation of a GFT quantum over the Fock vacuum, written as \begin{equation}
|g_I\rangle=\hat{\varphi}^{\dagger}(g_I)|\emptyset\rangle~,
\end{equation}
corresponds to the creation of a single open four-valent loop quantum gravity spin network vertex.
One can thus construct N-particle
states, which are needed to describe extended three-geometries, and then use second-quantised hermitian operators to obtain  quantum geometric observable data for these states. 

Group Field Cosmology (GFC)~\cite{GFC} is the quantum cosmological model based on GFT, in the sense that the condensate phase is identified with the continuum approximately smooth and spatially homogeneous space-time of our universe. In this context one applies mean field techniques for the corresponding GFT states and obtains an effective dynamics to which one attempts a cosmological interpretation. 
The geometric interpretation that one may have in mind is that of a gas of tetrahedra, each of them in the same microscopic quantum state, defined by the condensate mean field $\sigma(g_I)$ with
\begin{equation}
\langle\sigma|\hat{\varphi}|\sigma\rangle=\sigma\neq 0~.
\end{equation}
Separate the field operator into the condensate mean field $\sigma(g_I)$ and the non-condensate contribution $\delta\hat{\varphi}(g_I)$ as
\begin{equation}
\hat{\varphi}(g_I)= \sigma(g_I)+\delta\hat{\varphi}(g_I)~,
\end{equation}
and consider  the coherent (as eigenstate of the field operator) state
\begin{equation}
|\sigma\rangle=e^{-\frac{1}{2}\int(dg)^d|\sigma(g_I)|^2}~e^{\hat{\sigma}}|\emptyset\rangle~~\mbox{with}~~\hat{\sigma}=\int(dg)^d\sigma(g_I)\hat{\varphi}^{\dagger}(g_I)~,
\end{equation}
constructed from quantum tetrahedra which encode the same quantum geometric information.
The effective dynamics is obtained from the quantum cosmology equation 
\begin{equation}
\frac{\delta S[\sigma,\bar{\sigma}]}{\delta\bar{\sigma}(g_I)}=\int(dg')^d\mathcal{K}(g_I,g'_I)\sigma(g'_I)+\frac{\delta\mathcal{V}}{\delta\bar{\sigma}(g_I)}=0~,
\label{qc-eq}
\end{equation} 
with the action as in Eq.~(\ref{cl-ac}) but now written in terms of the condensate field $\sigma$, while the kinetic kernel is 
\begin{equation}
\mathcal{K}=\delta(g'_{I}g^{-1}_{I})\delta(\phi'-\phi)
\biggl[-(\tau\partial_{\phi}^2+\sum_{I=1}^4
\Delta_{g_I})+m^2\biggr],~~~\tau>0~,
\end{equation}
where $\phi$ is a massless scalar field, playing the r\^ole of a
relational clock~\cite{GFCEmergentFriedmann}.
Note that Eq.~(\ref{qc-eq}) is the analogue of Gross-Pitaevskii equation for real Bose condensates.

Consider first a static mean field $\sigma(g_I,\phi)=\sigma(g_I)$ and neglect all interactions setting ${\cal V}=0$ as in Ref.~\cite{GFClowspinStaticEffInt}. The corresponding equation, in an isotropic restriction, can be expressed in terms of the angle parametrisation $\psi$ as  
\begin{equation}
\label{eom}
-\Big[\frac{d^2}{d\psi^2}+2\cot(\psi)\frac{d}{d\psi}\Big]\sigma(\psi)+2\mu\sigma(\psi)=0~,~~~\psi\in\Big[0,\frac{\pi}{2}\Big]~,
\end{equation}
with $\mu\equiv m^2/12$.
Assuming as a boundary condition, the {\sl near-flatness} condition
\begin{equation}
\sigma(\psi)|_{\psi=\frac{\pi}{2}}=0~,
\end{equation}
the eigensolutions of Eq.~(\ref{eom}) are
\begin{equation}\label{solutionsangle}
\sigma_j(\psi)=\frac{\sin((2j+1)\psi)}{\sin(\psi)}~,~~~\psi\in\Big[0,\frac{\pi}{2}\Big]~,
\end{equation}
with $j\in \frac{2\mathbb{N}_0+1}{2}$ 
corresponding to the eigenvalues $\mu=-2j(j+1)$.
The probability density is concentrated around small $\psi$, namely around small curvature values~\cite{GFClowspinStaticEffInt}, implying that the building blocks of geometry are almost flat.
Studying the normalised spectrum of the loop quantum gravity volume operator
\begin{equation}
V=V_0\sum_{m\in\mathbb{N}_0/2} |\sigma_{j;m}|^2 V_m~~\textrm{with}~~V_m\sim m^{3/2}~,~V_0\sim l_{\rm Pl}^3~,
\end{equation}
(written in terms of Fourier coefficients $\sigma_{j;m}$)
with respect to the eigensolutions $\sigma_j(\psi)$, we find that eigensolutions for smaller $j$ have a bigger volume, while the volume remains finite for all $j$. Thus, a general solution of the quantum cosmology equation, decomposed in terms of eigensolutions, describes a finite space with the largest contributions from the low spin modes. The same conclusion is drawn studying the normalised spectrum of the area operator
\begin{equation}
A=A_0\sum_{m\in\mathbb{N}_0/2} |\sigma_{j;m}|^2 A_m~~\textrm{with}~~A_m\sim (m(m+1))^{1/2}~,~V_0\sim l_{\rm Pl}^2~,
\end{equation}
with respect to the eigensolutions $\sigma_j(\psi)$.

Let us now turn on the interaction term to study the impact of simplified interactions onto static GFT quantum gravity condensates which are thought to describe continuous three-geometries~\cite{GFClowspinStaticEffInt}. Consider the effective potential
\begin{equation}
V_{\rm eff}(\sigma)={\mu\over 2}\sigma^2 +{\kappa\over 4}\sigma^4~,
\end{equation}
with $\mu<0$ and $\kappa>0$ in order to get a nontrivial (nonperturbative) vacuum with nonzero vacuum expectation value, $\langle\sigma\rangle\neq 0$, so that there is agreement with the condensate state ansatz. The system would settle into one of the two minima $\langle\sigma_0\rangle=\pm\sqrt{-\mu/\kappa}$ and describe a condensate.
Considering the interaction term as a perturbation of the free case, one then analyses the effect of perturbations onto the spectra of geometric operators. Studying the probability density of the interacting mean field over $\psi$, we conclude that the finiteness of the free solutions at the origin is lost due to the interactions. However, the concentration of the probability densities around the origin can still be maintained, giving rise to nearly flat solutions, as long as $|\kappa|$ does not become too big. Studying the normalised spectrum of the volume and area operators 
with respect to the interacting mean field $\sigma_j$, we conclude that perturbations increase both the volume and the area, but remain finite in the weakly nonlinear regime; these effects are more pronounced for small values of $j$. We have shown the validity of these results for 
\begin{equation}
V_{\rm eff}(\sigma)={\mu\over 2}\sigma^2 +{\kappa\over 4}\sigma^5~,
\end{equation}
with either $(\mu<0, \kappa>0)$ or $(\mu<0, \kappa<0)$ in order to have a nontrivial nonperturbative vacuum.
To investigate the condensate phase we have also studied the solutions around the nontrival minima, and conclude that  the condensate consists of many discrete building blocks predominantly of the smallest nontrivial size. Summarising, a free condensate configuration in an isotropic restriction settles dynamically into a low-spin configuration of the quantum geometry. Hence, our analysis supports the proposal that an effectively continuous geometry may emerge from the collective behaviour of a discrete pre-geometric GFT substratum.

One may then lift the isotropic restriction to investigate anisotropic GFT condensate configurations. Studying the probability density of the mean field, we have found that anisotropies  play an important r\^ole only at small values of the relational clock (i.e., at small volumes), whereas at late times the isotropic mode dominates~\cite{Pithis:2016cxg}.

\subsection{Cosmological consequences through the effective Friedmann equation}
The effective action to describe the dynamics of a GFT condensate reads
\begin{equation}
S=\int d\phi(A |\partial_\phi \sigma|^2 +{\cal V}(\sigma))~,
\label{act}\end{equation}
where $\sigma$ is a complex scalar field representing the configuration of the Bose condensate of GFT quanta as a function of the massless scalar field $\phi$ that plays the r\^ole of relational time, and $\rho_j$ is the modulus of the component of $\sigma$ corresponding to the spin-$j$ representation of SU(2). For convenience,  we separate $\sigma_j$ into its modulus $\rho_j$ and phase $\theta_j$ as $\sigma(\phi)=\rho_j e^{i\theta_j}$. 

Assuming the interactions between GFT quanta as being sub-dominant, the corresponding equations for the modulus and phase of $\sigma_j$ read~\cite{GFClowspinStaticEffInt} 
\begin{eqnarray}
\rho_j''-[m_j^2+(\theta_j^\prime)^2]\rho_j &\approx& 0~,\nonumber\\
2\rho_j'\theta_j^\prime+\rho_j\theta_j''&\approx& 0~,
\end{eqnarray}
leading to two constant charges
\begin{eqnarray}
E_j&\approx&(\rho_j^\prime)^2 + \rho_j^2(\theta_j^\prime)^2-m_j^2\rho_j^2~,\nonumber\\
Q_j&\approx&\rho_j^2 \theta_j^\prime~;
\end{eqnarray}
the former called the {\sl GFT energy}.

The expectation values for the total volume and momentum of the massless scalar field operators with respect to $\phi$, evaluated on isotropic GFT condensate states read:
\begin{eqnarray}
V(\phi)&=&\sum_{j\in \mathbb{N}_0/2}V_j[\rho_j(\phi)]^2~,\nonumber\\
\pi_\phi(\phi)&=&{\hbar\over 2i}\sum_j\Big(\bar\sigma_j(\phi) \sigma_j(\phi)^\prime-\bar\sigma_j(\phi)^\prime \sigma_j(\phi)\Big)~.
\end{eqnarray}
Since $\partial\pi_\phi/\partial\phi=0$, which is the continuity equation on cosmology for the case of a massless scalar field, one concludes that the effective Friedmann and acceleration equations in the semi-classical limit read:
\begin{align}
\frac{\partial_{\phi}V}{V}&=2\frac{\partial_{\phi}\rho}{\rho}\equiv 2g(\phi)~,\label{eq:EmergingFriedmannI}\\ \frac{\partial^2_{\phi}V}{V}&=2\left[\frac{\partial^2_{\phi}\rho}{\rho}+\left(\frac{\partial_{\phi}\rho}{\rho}\right)^2\right]\label{eq:EmergingFriedmannII}~,
\end{align}
respectively. 
Note that in Eq.~(\ref{eq:EmergingFriedmannII}) we have omitted the index $j$ in $\rho_j$ and hence wrote $V(\phi)$ as $V(\phi)=V_j\rho^2(\phi)$, since as we have shown earlier, the condensate field peaks on a particular representation $j$.

The evolution equation of the universe in the classical limit of GFT can be written as an effective Friedmann equation:
\begin{equation}
H^2=\Big({\partial_\phi V\over 3V}\Big)^2\dot\phi^2\equiv {8\over 9}g^2\epsilon~,
\end{equation}
where $\epsilon$ plays the r\^ole of an energy density.
One can thus define an {\sl effective gravitational constant} $G_{\rm eff}=g^2/(3\pi)$, from the collective behaviour of space-time quanta. 

The proposal of an effective gravitational constant is not new, and as it has been shown in the literature, it may need to interesting consequences~\cite{deCesare:2016dnp}.
In particular, quantum geometry effects may lead to stochastic fluctuations of the gravitational constant, which can be thus considered as a macroscopic effective dynamical quantity. In consequence, a time-dependent dark energy term in the modified field equation can be expressed in terms of a time-dependent dynamical gravitational constant. As a result, the late-time accelerated expansion of the universe may be ascribed to quantum fluctuations in the geometry of space-time rather than the vacuum energy from the matter sector~\cite{deCesare:2016dnp}.
 
From Eq.~(\ref{eq:EmergingFriedmannI}) it is clear that there is a value of $\phi$, call it $\Phi$, for which $g(\phi)$ vanishes, and hence at relational time $\Phi$ there is a {\sl bounce}, where the energy density reaches a maximum value 
\begin{equation}
\epsilon_{\rm max}={1\over 2}{Q^2\over V^2_{\rm bounce}}~,
\end{equation}
with $V_{\rm bounce}$ the volume at the bounce:
\begin{equation}
V_{\rm bounce}={V_{j0}(\sqrt{E^2+12\pi GQ^2}-E)\over 6\pi G}~,
\end{equation}
which gives the minimum (non-zero) value of the volume.
Note that the singularity is always avoided for $E<0$ and provided $Q$ is non-zero, it is also avoided for $E>0$.

Next issue we would like to address is whether one can get an early era of geometric-type inflation, meaning an early era of accelerated expansion through the different geometric background and not via the introduction of an inflaton field with an {\sl ad hoc} potential. In standard (classical) cosmology, there is a scale factor $a$, leading to a volume $V=a^3$ and a {\sl proper time} $t$. In the GFT context there is no metric, neither proper time, nevertheless we will be able to define an equivalent to $d^2 a/dt^2>0$ condition, via the volume $V$. More precisely, within the cosmological context derived from GFT, we have
\begin{equation}
{\ddot a\over a} ={1\over 3} \Big({\pi_\phi\over V}\Big)^2\Big[{\partial^2_\phi V\over V} -{5\over 3} \Big({\partial_\phi V\over V}  \Big)^2  \Big]~,
\end{equation} 
leading to the following condition for inflation
\begin{equation}
{V^{''}\over V}>{5\over 3}\Big({V'\over V}  \Big)^2~,
\end{equation}
which is valid even in the absence of space-time and the absence of proper time.
Hence, within the GFT condensate cosmology, the condition for inflation reads~\cite{mm}
\begin{equation}
4m^2+{2E\over \rho^2}>{20\over 3}g^2~,
\end{equation}
and near the bounce the l.h.s. of the above equation is positive while the r.h.s. is zero. In conclusion, there is an early era of geometric inflation, namely an accelerated expansion in the absence of an inflaton field, for either $E<0$ or $E>0$; in the latter case the accelerated expansion is followed by a maximal deceleration~\cite{mm}.

Certainly the next question is to study whether such a geometric inflation can last sufficiently long, therefore one has to calculate the number of e-folds $N$ defined here as
\begin{equation}
N={2\over 3}\log\Big({\rho_{\rm end}\over \rho_{\rm bounce}  }  \Big)~.
\end{equation}
A careful analysis shows that in this case, where interactions have been neglected, the number of e-folds is $0.119\leq N\leq 0.186$ for all possible values of $m^2$ and $Q^2$. Thus, GFT cosmology in the absence of interactions between building blocks cannot replace the standard inflationary scenario. In the following we will turn on the interactions, which is indeed the more natural and consistent scenario, as the quanta of geometry must be somehow glued with each other. Depending on our results we may also be able to give a prescription for the way one must build the GFT model and assign specific type of interactions in order to get at the semi-classical limit the desired properties for the space-time that must be then seen as the emergent 3-geometry of our homogeneous and isotropic universe~\cite{mam}.

Let us consider the effective action (\ref{act}) for an isotropic GFT condensate, and take~\cite{mam}
\begin{equation}
{\cal V}(\sigma)=B|\sigma(\phi)|^2+{2\over n} w|\sigma|^n+{2\over n^\prime} w^\prime|\sigma|^{n^\prime}~, \mbox{with}~~~n^\prime>n~.
\label{potential}\end{equation} 
Such a choice is motivated from {\sl spin foam} models for four-dimensional quantum gravity, which are mostly based on interaction terms of power 5 of the modulus field, as well as from {\sl tensor} models, which are based on even powers of $\sigma$; the former are called {\sl simplicial} and the latter {\sl tensorial}.

A detailed study has shown~\cite{mam} that the occurrence of a bounce and an early epoch of accelerated expansion found in the free theory, both hold in the interacting case. In the latter case, the solutions are cyclic motions describing oscillations around a stable minimum, hence interactions induce recollapse leading to cyclic cosmologies. Furthermore, in the interacting case, one can find the range of the parameter space for which inflation will last sufficiently long. Note that to have a successful inflationary scenario one will need to impose an extra condition, namely that there is no intermediate stage of deceleration between the bounce and end of inflation. Hence, one concludes that for an interacting potential, as in (\ref{potential}), with
\begin{equation}
\lambda<0 ~~ \mbox{and} ~~n\geq 5 ~ (n^\prime>n)~,\nonumber
\end{equation}
GFT cosmology can lead to an inflation-like era.

At this points let me make some remarks about the validity of the above results for a GFT model based on a real field.
As analysed in Ref.~\cite{Pithis:2016cxg}, for real-valued condensate fields, solutions which avoid the singularity problem and grow exponentially after the bounce can only be found for negative GFT energy $E$. Studying the stability properties of an evolving isotropic system, it was shwon that the condensate models give rise to effectively continuous, homogeneous and isotropic three-geometries built from many smallest and almost flat building blocks of the quantum geometry. Finally, studying the stability properties of an evolving anisotropic system, it was shown the isotropisation for increasing values of the clock (i.e., the volume), and that the singularity avoidance is not altered by the occurrence of anisotropies.

\section{The gravitational sector of Noncommutative Spectral Geometry}
Here, one follows the approach that in order to construct a quantum theory of gravity coupled to matter, the gravity-matter interaction is the most important ingredient to determine the dynamics. The aim will be to guess the small-scale structure of space-time from our knowledge at the electro-weak scale, which is well-studied by collider experiments. We consider the  Standard Model (SM) of particle physics, the most successful particle physics model we have at hand, as a phenomenological model which dictates the space-time geometry. Close but below the Planck scale, gravity and the SM
fields are thought to be packaged into geometry and matter on a Kaluza-Klein non-commutative space-time,  a four-dimensional space-time with an internal discrete zero dimensionality-space attached to each point.
Thus, space is defined as the product
of a four-dimensional compact Riemannian spin manifold ${\cal M}$ with
an internal Kaluza-Klein space ${\cal F}$.  Such simple non-commutative spaces ${\cal M} \times {\cal F}$, where
the non-commutative algebra describing space is the algebra of functions over ordinary space-time,
are called {\sl almost commutative manifolds}~\cite{ncg-book}.

The almost commutative manifold  ${\cal M} \times {\cal F}$ is defined by the canonical spectral triple $({\cal A}, {\cal H}, D)$, in terms of an algebra, a Hilbert space  and a Dirac operator, as~\cite{ccm}:
\begin{equation}
{\cal M}\times{\cal F}:=(C^\infty({\cal M},{\cal A_F}), L^2({\cal M},
S)\otimes {\cal H_F}, \slashed {\cal
  D}\otimes\mathbb{I}+\gamma_5\otimes {\cal D}_{\cal F})~;
\nonumber
\end{equation}
the set $C^\infty({\cal M})$ of square-integrable spinors $S$ on ${\cal
  M}$ forms  an algebra under point-wise multiplication acting on the Hilbert space $L^2({\cal M}, S)$ as multiplication operators, $\slashed {\cal D}$ is
the Dirac operator $-i\gamma^\mu\nabla_\mu^s$ acting as first order
differential operator on the spinors, the operator $\gamma_5$ plays the r\^ole of a chirality operator. In addition one considers the antilinear
isomorphism $J_{\cal M}$ as the charge conjugation operator on spinors.
The internal discrete space ${\cal F}$ is defined by the spectral triple
$({\cal A}_{\cal F}, {\cal H}_{\cal F}, D_{\cal F})$; it encodes the internal degrees of freedom at each point in space-time. The Hilbert space ${\cal H}_{\cal F}$ is a $96\times 96$ matrix, written in terms of $3\times 3$ Yukawa mixing matrices and a real constant responsible for neutrino mass terms. To recover the SM, the algebra ${\cal A_F}$ must be appropriately chosen. It was shown that~\cite{Chamseddine:2007ia}
\begin{equation}
{\cal A_F}=M_a(\mathbb{H})\oplus M_k(\mathbb{C})~,
\nonumber
\end{equation}
with $k$ an even number $k=2a$. The choice $k=4$ is the smallest possible value that produces correctly  16 fermions
 in each of the three generations.

It is important to note the physical meaning of the choice of such a non-commutative manifold. The spectral triple 
\begin{equation}
({\cal A, H, D}, J, \gamma)=(C^\infty({\cal M}), L^2({\cal
     M},S),\slashed{\partial}_{\cal M}, J_{\cal
     M},\gamma_5)\otimes({\cal A_F, H_F, D_F}, J_{\cal F},\gamma_{\cal
     F})~, \nonumber
\end{equation}
defining ${\cal M}\times {\cal F}$, is the product 
\begin{equation}
({\cal A}_1, {\cal H}_1, {\cal D}_1, J_1,
  \gamma_1)\otimes({\cal A}_2, {\cal H}_2, {\cal D}_2, J_2,
  \gamma_2)~, \nonumber\\
\end{equation}
with
\begin{equation}
{\cal A}={\cal A}_1\otimes{\cal A}_2~,~{\cal H}={\cal H}_1\otimes{\cal H}_2
~,~ {\cal D}={\cal D}_1\otimes 1 +\gamma_1\otimes{\cal D}_2~,
\gamma=\gamma_1\otimes\gamma_2~,~J=J_1\otimes J_2~.
\nonumber
\end{equation}
This {\sl doubling} of the algebra is essential in order to lead to the SM physics. As we have shown in Refs.~\cite{PRD,mvpm} this doubling is intimately related to dissipation,
gauge field structure, neutrino mixing, while it incorporates the seeds of
quantisation. 

To explore another physical meaning of the  almost-commutative manifold, let us identify it with a pair of four-dimensional Minkowski space-times
embedded in a five-dimensional Minkowski geometry. Following this approach,  we have considered~\cite{km2} fermions travelling within
the light cone of the ambient five-dimensional space-time, and then derived the energy-momentum
dispersion relation, thus arguing for its non-commutative geometry origin.

To write down the action we will use the spectral action principle, according which  
the bosonic action functional depends only on the
spectrum of the fluctuated Dirac operator ${\cal D}_A$ (a Dirac operator with a gauge connection):
\begin{equation}
{\cal D}_A={\cal D} + A + \epsilon' J A J^{-1}~,
\end{equation}
with $A$ a self-adjoint operator,
$J$ an anti-unitary operator and $\epsilon\in\{\pm 1\}$,
and is of the form
\begin{equation}
{\rm Tr}(f({\cal D}^2_A/\Lambda^2))~,
\label{bosonic}
\end{equation}
with $f$ a cut-off function and $\Lambda$ the cut-off energy scale at
which the action holds. This action sums up all eigenvalues of the fluctuated
Dirac operator ${\cal D}_A$ which are smaller than the cut-off energy
scale $\Lambda$. We evaluate the trace using heat kernel
techniques, in terms of  the Seeley-de Witt coefficients
$a_n$; its asymptotic expansion reads
\begin{equation}
\label{as-exp}
{\rm Tr}(f({\cal D}^2_A/\Lambda^2))\sim 2f_4\Lambda^4 a_0({\cal
  D}_A^2) + 2f_2\Lambda^2 a_2({\cal D}_A^2) + f(0)a_4({\cal D}_A^2)
+{\cal O}(\Lambda^{-2})~.
\end{equation}
Hence, the cut-off function $f$ plays a r\^ole through only three of its momenta, $f_4, f_2, f_0$, which are respectively
related to the cosmological constant, the gravitational constant and
the coupling constants at unification.  
One must then add the fermionic part of the action:
\begin{equation}
(1/2)\langle J\Psi, {\cal D}_A \Psi\rangle~;~ \Psi\in{\cal H}^+~.
\end{equation}
After a long calculation, the bosonic part of the spectral action can be schematically written as~\cite{ccm}
\begin{eqnarray}
S_{\Lambda} &=& \int d^4x \sqrt{g} \left( A_1 \Lambda^4 + A_2 \Lambda^2\left(\frac 54 R - 2 y_{\rm t}^2H^2 - M^2 \right)\right.  \nonumber\\&& \ \ \ \ \ \ \ \ \ \ \ \ \ \ +A_3\left( g_2^2W^{\alpha}_{\mu\nu}W^{\alpha~\mu\nu} + g_3^2 G^a_{\mu\nu}G^{a~\mu\nu} + \frac{5}{3}g_1^2 B_{\mu\nu}B^{\mu\nu}\right)\nonumber\\&&  \ \ \ \ \ \ \ \ \ \ \ \ \ \ + \mbox{other ${\cal O}(\Lambda^0)$ }
\left. 
+ \mbox{other} \ {\cal O}\left(\Lambda^{-2}\right) \right)~, \label{sa}
\end{eqnarray}
where $y_{\rm t}$ is the Yukawa coupling for the top quark; $R$ is the curvature scalar; $W,G,B$ are curvature tensors for the three interactions and $M$ is (up to a numerical factor)  a heavy Majorana right-handed neutrino mass. The $A_i$ with $i=1,2,3$ are constants which depend on the details of the cut-off function $f$ and for typical choices of that function the three constants are not too different from unity.
Finally, the  $g_1$, $g_2$, $g_3$ are the corresponding gauge couplings of the three interactions; they satisfy
$g_2^2=g_3^2=(5/3)g_1^2$, which hold in several Grand Unified
Theories (GUTs), hence the cut-off scale $\Lambda$ can be considered 
as the GUT scale, $\Lambda \sim(10^{14}-10^{17})$~GeV. 

It is worth noting that very few gauge groups fit into the spectral action model. The standard model does as well as the Pati-Salam group SU(2)$_L\times$ SU(2)$_R\times$ SU(4), but SO(10), of which the former group is an intermediate breaking stage, does not. The absence of larger groups with novel representation is a positive outcome since it prevents proton decay.

Using renormalisation group techniques one then derives predictions for the SM phenomenology; the results are in agreement with current particle physics data, including the Higgs mass~\cite{pheno}.

The gravitational sector of the theory coupled to matter, in Euclidean signature, reads~\cite{ccm}
\begin{eqnarray}\label{bsa}
{\cal S}_{\Lambda}= \int \left(
\frac{1}{2\kappa_0^2} R + \alpha_0
C_{\mu\nu\rho\sigma}C^{\mu\nu\rho\sigma} + \gamma_0 +\tau_0 R^\star
R^\star
\right.  
+ \frac{1}{4}G^i_{\mu\nu}G^{\mu\nu
  i}+\frac{1}{4}F^\alpha_{\mu\nu}F^{\mu\nu\alpha}
\nonumber\\ 
+\frac{1}{4}B^{\mu\nu}B_{\mu\nu}
+\frac{1}{2}|D_\mu{\bf H}|^2-\mu_0^2|{\bf H}|^2
\left.
- \xi_0 R|{\bf H}|^2 +\lambda_0|{\bf H}|^4
\right) \sqrt{g} \ d^4 x~,
\end{eqnarray}
where 
\begin{eqnarray}
\kappa_0^2=\frac{12\pi^2}{96f_2\Lambda^2-f_0\mathfrak{c}}~&,&~
\alpha_0=-\frac{3f_0}{10\pi^2}~,\nonumber\\
\gamma_0=\frac{1}{\pi^2}\left(48f_4\Lambda^4-f_2\Lambda^2\mathfrak{c}
+\frac{f_0}{4}\mathfrak{d}\right)~&,&~
\tau_0=\frac{11f_0}{60\pi^2}~,\nonumber\\
\mu_0^2=2\Lambda^2\frac{f_2}{f_0}-{\frac{\mathfrak{e}}{\mathfrak{a}}}~&,&~
\xi_0=\frac{1}{12}~,\nonumber\\
\lambda_0=\frac{\pi^2\mathfrak{b}}{2f_0\mathfrak{a}^2}~&,&~
{\bf H}=(\sqrt{af_0}/\pi)\phi~; 
\end{eqnarray}
${\bf H}$ a rescaling of the Higgs field $\phi$ to normalise the
kinetic energy, and the momentum $f_0$ is physically related to the
coupling constants at unification. 

The action (\ref{bsa}) is obtained from the {\sl cut-off} bosonic spectral action (\ref{bosonic}). It 
 contains the Einstein-Hilbert action with a cosmological term, a topological term, a conformal gravity term with the Weyl curvature tensor, as well as conformal coupling of the Higgs field to gravity. The coefficients of the gravitational terms depend upon the Yukawa parameters of the particle physics content. Given that the cut-off scale is  $\Lambda \sim(10^{14}-10^{17})$~GeV, the action (\ref{bsa}) provides a natural framework to address early universe cosmology.

Since this action  contains higher derivative terms, one may wonder whether this gravitational theory
is plagued by linear instabilities, namely by the appearance of negative energy modes. We have shown~\cite{km}  that there
exists a class of almost commutative torsion geometry that leads to a Hamiltonian which is bounded from below and
hence argued that the theory does not suffer from linear instabilities.

Studying the low-energy weak curvature regime, where one may neglect the non-minimal coupling term between the background geometry and the Higgs field, we have shown that any modifications to the background equation may appear
at leading order only for anisotropic and inhomogeneous models~\cite{Nelson:2008uy}.
As energies are approaching the Higgs scale, the non-minimal coupling
of the Higgs field to the curvature can no longer be neglected. In the latter case,  the coupling plays the r\^ole of an effective gravitational constant; alternatively, the coupling increases the self-interaction of the Higgs field.
Note that one observes a similarity with compactified string models and
with chameleon cosmology.

Exploring the cosmological r\^ole of the Higgs field~\cite{nm,mmm}, which has conformal coupling to gravity, we have calculated the renormalisation of the Higgs self-coupling up to
two-loops and shown that while the Higgs potential can lead to slow-roll
inflation, the CMB constraints turn the predictions of such a scenario incompatible with the measured
value of the top quark mass~\cite{mmm}.

The gravitational sector of the bosonic spectral action may be seen as an
extended theory of gravity. Studying its astrophysical consequences  one may constrain $f_0$ (or equivalently $a_0=
-3f_0/(10\pi^2)$), one of the three free parameters.

Consider linear perturbations around Minkowski background~\cite{Nelson:2010rt} and study the propagation of gravitational waves. Comparing with data from 
binary pulsar systems, for which the rate of change of the
orbital frequency is well-known, one finds~\cite{Nelson:2010ru}
\begin{equation}
\beta \gtrsim 7.55\times 10^{-13} {\rm m}^{-1}~.\nonumber
\end{equation}
This constraint is improved using Gravity Probe B~\cite{gam}:
\begin{equation}
\beta\gtrsim 7.1\times 10^{-5} {\rm
   m}^{-1}~,
\nonumber
\end{equation}
or LARES data~\cite{sgaam}:
\begin{equation} 
\beta\gtrsim 1.2\times 10^{-6} {\rm m}^{-1}~.
\end{equation}
However, the strongest constraint on $\beta$ is obtained
through torsion balance experiments~\cite{gam}:
\begin{equation}
\beta\gtrsim  10^4 {\rm m}^{-1}~.
\end{equation}

To obtain the cut-off bosonic spectral action we have made an expansion in $1/\Lambda^2$. This expansion is however valid when the fields and their derivatives are small with respect to the cut-off scale $\Lambda$, hence the spectral action (\ref{as-exp}) is valid only in the weak-field approximation.
To cure this problem, we have proposed a novel definition of the bosonic spectral action using zeta function regularisation:
\begin{equation}
S_\zeta \equiv \lim_{s\rightarrow 0} \Tr D^{-2s}\equiv \zeta(0,D^2)~. \label{szetadef}
\end{equation}
The $\zeta$ function is well defined and given by the $a_4$ 
heat kernel coefficient associated with the Laplace type operator $D^2$:
\begin{eqnarray}
S_\zeta &=& a_4\left[D^2\right] = \int d^4 x \,\sqrt{g}\,L~,\nonumber\\
\mbox{with} \quad L(x) &=& a_4(D^2,x)~. \label{BSAv20}
\end{eqnarray}
Following this approach one can address issues of renormalisability and spectral dimensions. More precisely,
in the zeta spectral action there are no operators
of dimension higher than four, hence the theory is renormalisable. The zeta spectral action can be used up to the Planck scale where  the very nature of space-time changes due to quantum gravitational effects. Moreover, the $\zeta$ spectral action  exhibits viable spectral dimensions.  More precisely, for Higgs scalars and gauge bosons the spectral dimensions coincide with the topological one and equal four, while in the gravitational sector the spectral dimension equals two, which implies improved ultraviolet behavior of the gravitational propagators.

\section{Conclusions}
To address remaining open questions in early universe cosmology we need a quantum gravity theory, while in return cosmology is the best (if not the only) way to test the validity of quantum gravity proposals. I have highlighted three cosmological models built upon three quantum gravity proposals from string/brane theory, group field theory and non-commutative geometry. I have shown how some issues may be resolved, in particular regarding an early era of accelerated expansion, growth of matter perturbations and resolution of the big bang singularity.
I have then argued how one can get a purely geometric explanation for the standard model of particle physics, which still remains the best particle physics model we have at hand.

Given these promising results, one may further explore predictions of quantum gravity models which will allow to adjust their free parameters, modify them or even disprove them. A synergy between such proposals is required. For instance, one may investigate whether there is a semi-classical regime for which the effective dynamics of GFT can reduce to the dynamics of another non-perturbative approach to quantum gravity, like loop quantum cosmology. In addition, one may investigate the connection between non-perturbative approaches to quantum gravity, where the pre-geometric degrees of freedom are algebraic in nature,  and non-commutative spectral geometry,  which may help to quantise the latter.

\end{document}